\documentclass[aps,pra,twocolumn,groupedaddress,showpacs,10pt]{revtex4-1}
\usepackage{graphicx}
\usepackage{amsmath}
\usepackage{eurosym}
\usepackage{amsfonts}
\usepackage{color}

\begin{document}


\title{Spin noise spectroscopy under resonant optical probing conditions: coherent and non-linear effects}



\author{H. Horn}
\author{G.~M. M{\"u}ller}
\affiliation{Institute for Solid State Physics, Leibniz Universit{\"a}t Hannover, Appelstr. 2, 30167 Hannover, Germany}

\author{E. M. Rasel}
\affiliation{Institute for Quantum Optics, Leibniz Universit{\"a}t Hannover, Welfengarten 1, 30167 Hannover, Germany}

\author{L. Santos}
\affiliation{Institute for Theoretical Physics, Leibniz Universit{\"a}t Hannover, Appelstr. 2, 30167 Hannover, Germany}

\author{J. H{\"u}bner}
\author{M. Oestreich}
\affiliation{Institute for Solid State Physics, Leibniz Universit{\"a}t Hannover, Appelstr. 2, 30167 Hannover, Germany}

\date{\today}



\date{\today}

\begin{abstract}
High sensitivity Faraday rotation spectroscopy is used to measure the fluctuating magnetization noise of non-interacting rubidium atoms under resonant and non-resonant optical probing conditions. The spin noise frequency spectra in dependence on the probe light detuning with respect to the D2-transition reveals clear signatures of a coherent coupling of the participating electronic levels. The results are explained by extended Bloch equations including homogeneous and inhomogeneous broadening mechanisms. Our measurements further indicate that spin noise originating from excited states are governed at high intensities by collective effects.
\end{abstract}

\pacs{42.25.Bs, 42.50.Gy, 42.50.Lc, 32.30.-r}

\maketitle

\section{Introduction}
The concept of spin noise was first put forward by Bloch \cite{bloch:pr:70:460:1946}  as the incomplete stochastic cancellation of up- and down-spins in the case of a nuclear spin system. This nuclear spin noise was later experimentally verified by Sleator \textit{et al.} \cite{sleator:prl:55:1742:1985}.  Analogously to nuclear spins, the total spin of a classical gas of unpolarized atoms also fluctuates around a zero mean polarization. This atomic spin noise can be mapped onto the intensity of a probe laser due to dichroic bleaching of the relevant optical transitions \cite{AleksandrovJETP1977, McIntyreOL1993, takahisa:prl:84:5292:2000, MihailaPRA2006}. Ultimatively, Aleksandrov and Zapasskii demonstrated that the spin noise of an alkali gas can also be recorded in a perturbation-free manner by off-resonant Faraday rotation \cite{aleksandrov:jetp:54:64:1981}. For sufficient detuning from the probed optical resonance, the off-resonant probing \cite{happer:prl:18:577:1967} even establishes a quantum non-demolition measurement of the atomic spin \cite{kuzmich:pra:60:2346:1999} and represents the building block of several light-matter and matter-matter entanglement experiments \cite{julsgaard:nat:413:400:2001, julsgaard:nature:432:482:2004, sherson:nat:443:557:2006}. The relative  noise signal is large in small spin ensembles  \cite{crooker:nature:431:49:2004} which makes spin noise spectroscopy especially interesting for studying spin dynamics in semiconductor nanostructures \cite{oestreich:prl:95:216603:2005, mueller:physe:2010}. In semiconductors, optical excitations can fundamentally change the observed electron spin dynamics such that only spin noise spectroscopy employing below band gap light enables access to the intrinsic spin dynamics in these systems \cite{muller:prl:101:206601:2008}. Correspondingly, experimental as well as theoretical studies of spin noise spectroscopy in the case of atom as well as semiconductor physics usually focus onto the off-resonant probing regime and the weak scattering limit \cite{katsoprinakis:pra:75:042502:2007, TakeuchiPRA2007}. However, in the case of a strong inhomogeneously broadened optical transition, as present in an ensemble of self-ensembled semiconductor quantum dots, donor bound carriers, or a hot gas of atoms, a significant fraction of the contributing spins is inevitably resonantly probed by spin noise spectroscopy  \cite{crooker:prl:104:036601:2010, PhysRevA.83.032512}. 

In the following,  we present  an experimental and theoretical assessment of spin noise  under resonant as well as off-resonant probing conditions to reveal effects that accompany spin noise spectroscopy under resonant excitation. We chose a classical rubidium gas as an archetypal sample system and the transmitted laser light acts simultaneously as both, probing and pumping beam, respectively. The experiments reveal a pronounced non-linear behavior at the optical resonance \cite{BudkerRMP2002, AlexandrovJOSAB2005, MartinelliPRA2004}. The observations are well reproduced by extended Bloch equations where the coherent coupling of the involved atomic states is crucial to explain the experimental results. Experiment and calculations reveal that the observed spin noise under resonant probing differs qualitatively depending on whether the probed optical transition is dominantly inhomogeneously or homogeneously broadened. In the case of an inhomogeneously broadening, we furthermore encounter significant spin noise contributions from excited states. The complex experimental results can only partly be reproduced by a single-particle Bloch analysis which indicates the presence of a many-body effect such as nonlinear self-rotation and polarization selective absorption of the probe light.
\begin{figure}
\includegraphics[keepaspectratio, width=\columnwidth]{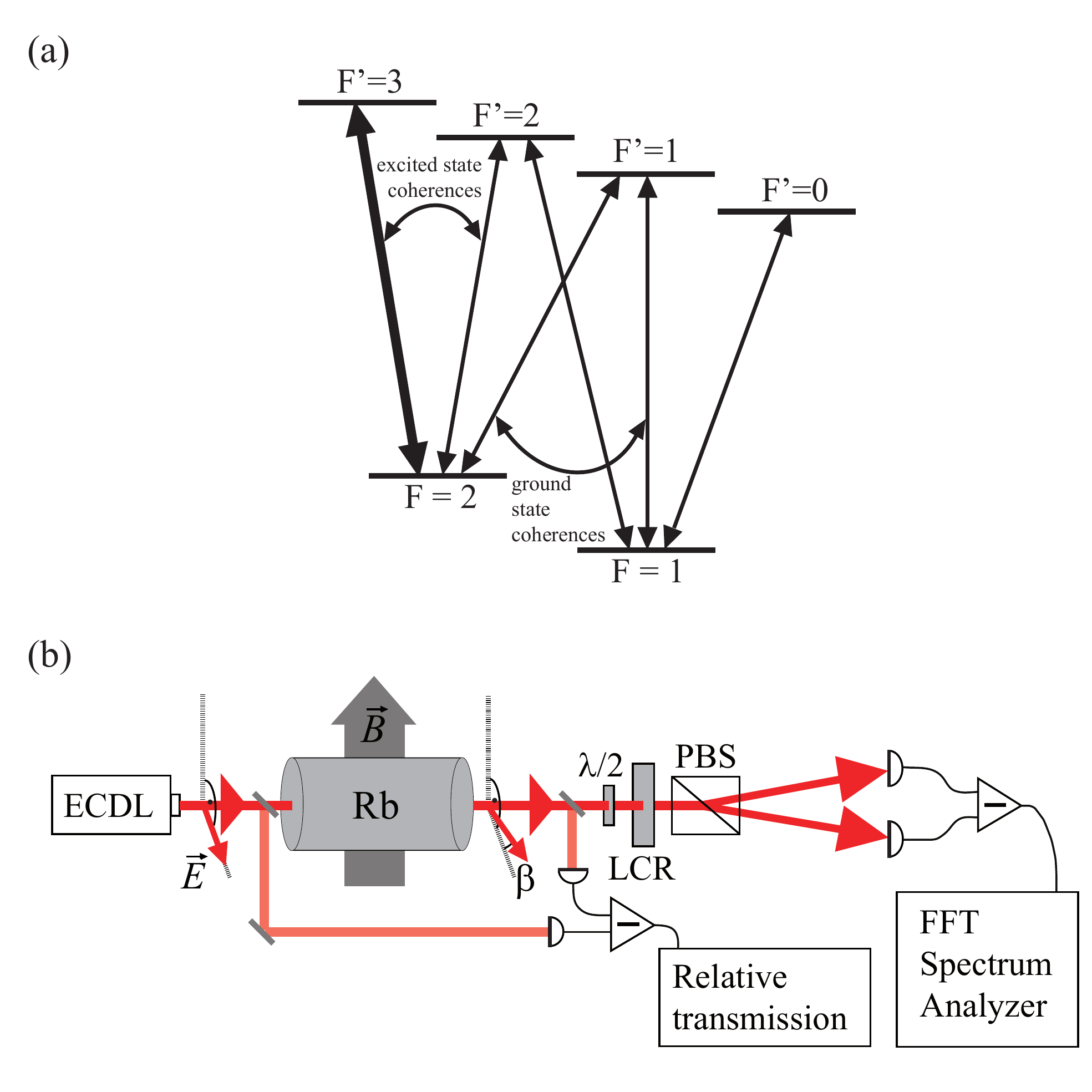}
\caption{(Color online) (a) D2 transition ($^2S_{1/2}\rightarrow {^2P}_{3/2}$) for the case of $^{87}$Rb. The bold arrow indicates the transition that corresponds to zero detuning in the following. Neighboring ground and excited states are coherently coupled. (b) Experimental spin noise set-up. A liquid crystal retarder (LCR) switches the polarization analysis between sensitivity on linear or circular birefringence, respectively (details are found in the text).}
\label{aufbau}
\end{figure}

An external magnetic field applied in transverse, i.e., Voigt, geometry leads to a splitting of the linear polarized $\sigma_{(at)}$ and $\pi_{(at)}$ atomic transitions (\textit{at}) in longitudinal excitation direction. The classical, dichroic Voigt effect could be observed if both transitions are addressed by linear polarized light, oriented $45^{\circ}$ with respect to the magnetic field axis. On the other hand, a pure Faraday rotation $\theta_F$ of linear polarized light occurs even at zero magnetic field if the refractive indices $n_{\pm}$ for the $\sigma^+$ and $\sigma^-$ polarization components differ due to dispersive effects \cite{YamamotoJOSA1979, MurooJOSAB1994}. Spin noise spectroscopy measures the stochastic spin polarization of the investigated system projected along the direction of light propagation via the Faraday effect. The spin noise arises due to the finite spin dephasing time $\tau_s$ and stochastic reorientation of the ensemble spin. A magnetic field transverse to the direction of light propagation modulates the spin noise with the Larmor frequency $\nu_{\mathrm{L}}=h^{-1} g_F\mu_{\mathrm{B}}B$ ($B$: magnetic field, $g_F$: Land\'e factor depending on the total angular momentum quantum number $F$, $\mu_{\mathrm{B}}$: Bohr magneton) and, correspondingly, facilitates spin noise measurements in the inevitable presence of white photon shot noise and common $1/f$ noise. The power spectrum of the measured Faraday rotation noise  is --- in the case of mono-exponential spin dephasing and a single, purely homogeneously broadened optical transition --- described by a Lorentz function centered at the corresponding Larmor frequency and a full width at half maximum (FWHM) given by $\nu_{\text{FWHM}}=(\pi\tau_s)^{-1}$ \cite{RomerRSI2007}. The detected Faraday rotation angle is proportional to (\textit{i}) the degree of spin polarization and (\textit{ii}) the circular birefringence given by the real part of the refractive index of the specimen \cite{mueller:physe:2010}.  (\textit{i}) The time averaged spin polarization vanishes at thermal equilibrium but has a finite mean deviation. The mean deviation of the stochastic spin polarization follows a Gaussian probability distribution.  (\textit{ii}) The investigated optical transition in this work is the Rubidium D2 $5^2S_{1/2}\rightarrow {5^2P}_{3/2}$ transition
which is exemplarily depicted in Fig.~\ref{aufbau}~(a) for $^{87}$Rb including the hyperfine splitting.  In off-resonant spin noise experiments, it usually suffices to model the probed optical transition by a single Lorentz oscillator. 
However, the complex details of the optical transition are needed to understand spin noise spectroscopy under resonant or nearly resonant probing conditions. The Faraday rotation of the linear light polarization results from circular birefringence, i.e., different refractive indices for the two circular polarization components which compile to linear polarized light. The corresponding dipole selection rules for circular polarized light demand $\Delta F =0,\pm 1$ and $\Delta m_F=\pm 1$. Figure \ref{aufbau}~(a) illustrates that resonant probing of the D2 transition evokes  coherences of the $S$ ground state as well as the excited $P$ states. 
We show in the following that the detailed structure of the D2 transition becomes evident as pronounced non-linearities in the spin noise signal under resonant excitation and 
a non-symmetric response for positive and negative detuning, respectively. We first present the experimental results (Sec.~\ref{Experiment}) and subsequently reproduce the spin noise spectra by a theoretical model that is based on the calculation of the refractive index of the atomic system by extended Bloch equations (Sec.~\ref{theory}).

\section{Experiment}\label{Experiment}
\paragraph*{Experimental technique.}
\begin{figure}
        \includegraphics[keepaspectratio, width=\columnwidth]{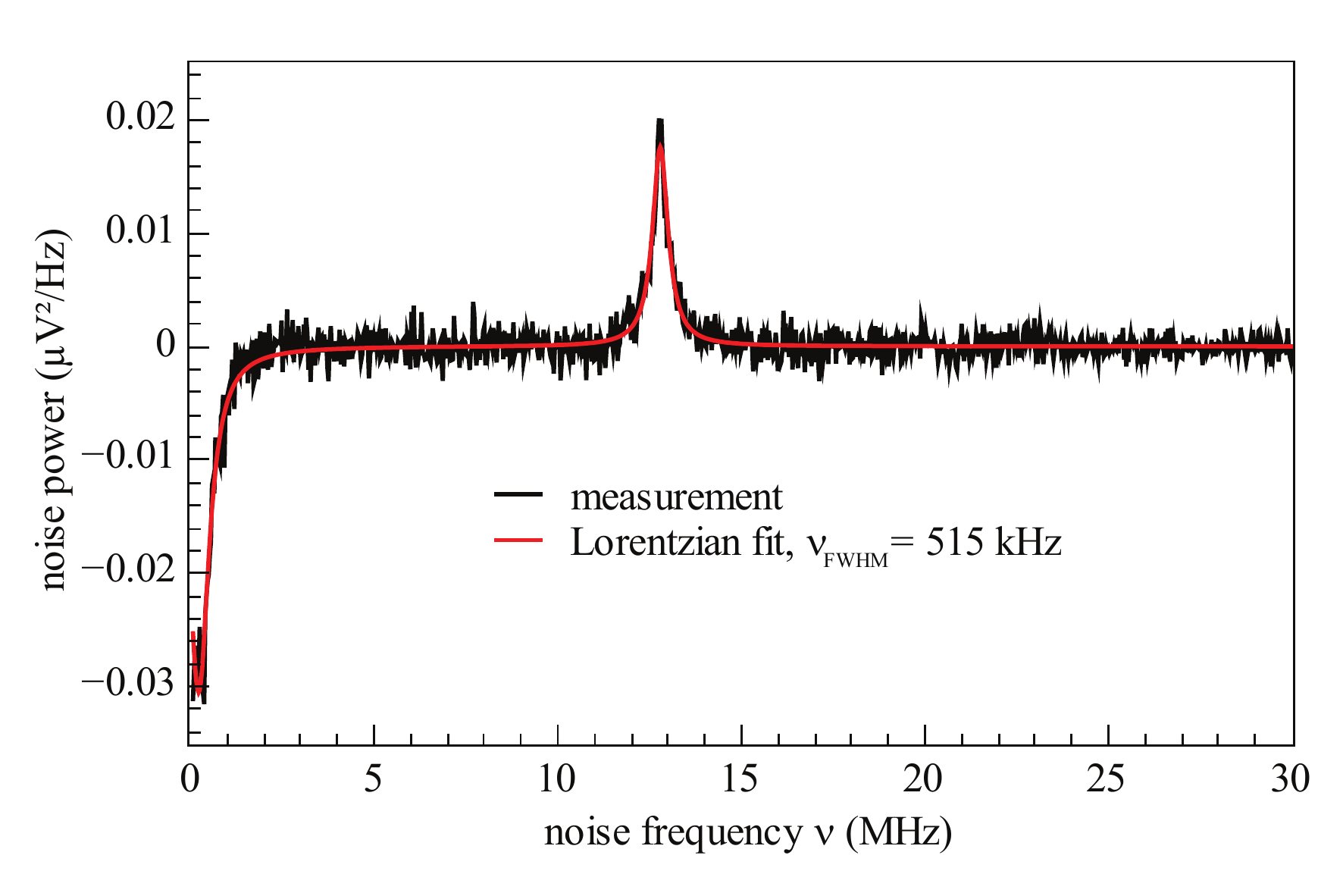}
        \caption{Off-resonant spin noise difference spectrum measured at $1.8\,\mathrm{mT}$ and $45\,\mu\mathrm{T}$ (terrestrial magnetic field), respectively.}
        \label{example}
\end{figure}
\begin{figure*}[ht]
        \includegraphics[width=\textwidth]{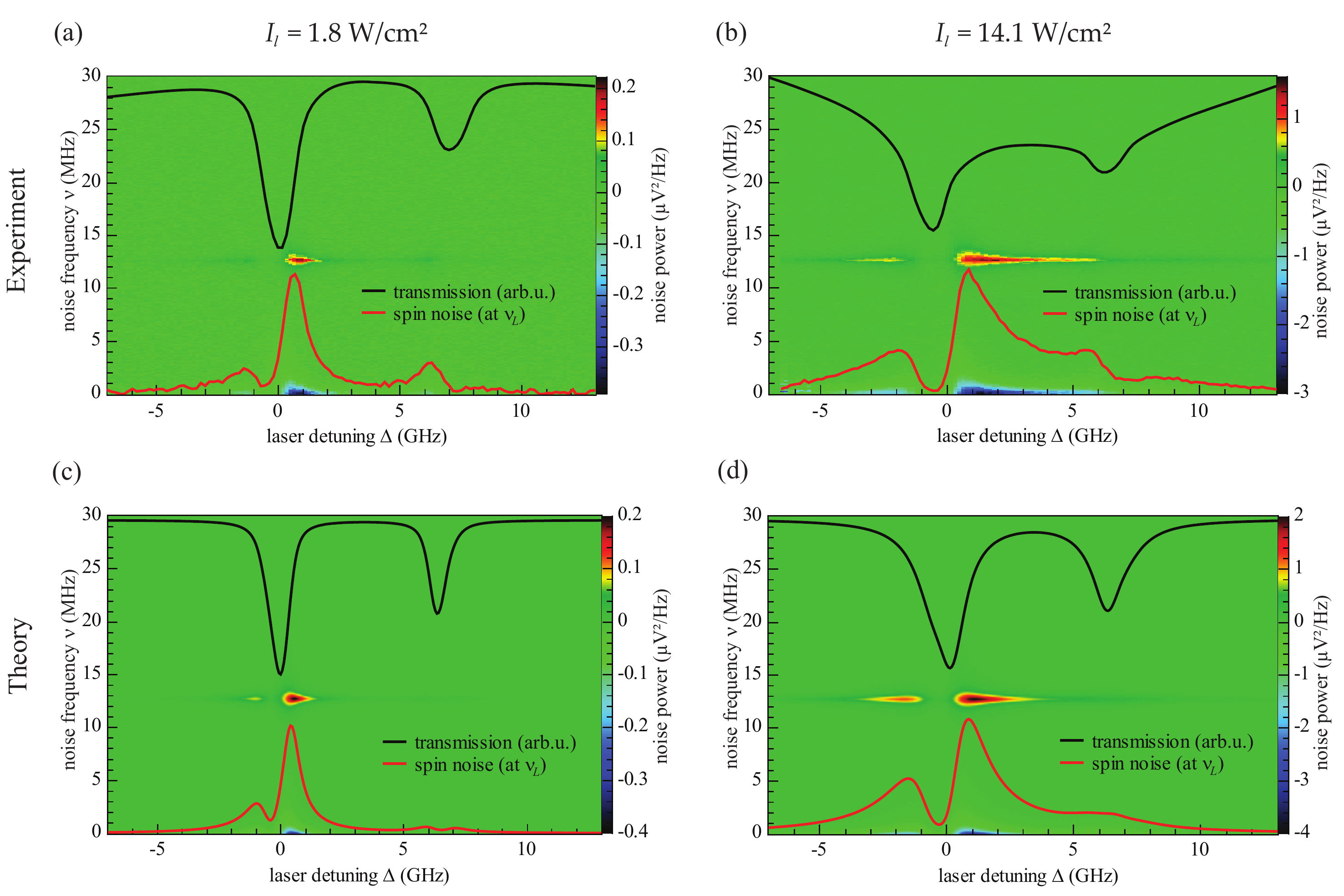}
        \caption{(Color online) Spin noise power density  in color coding as a function of laser detuning $\Delta$ (abscissa) and noise frequency $\nu$ (ordinate) for cell A (pure $^{87}$Rb). The black curve shows the absorption spectrum and the red curve the integrated noise power neglecting spin noise at noise frequencies centered at $0\,\mathrm{MHz}$. Zero detuning corresponds to the $F=2\rightarrow F'=3$ transition of $^{87}$Rb. (a,b) Experimental data with probe laser intensities $I_{l}=1.8$ and $I_{l}=14.1 \,\mathrm{W/cm^2}$. (c, d) Calculations corresponding to the experimental data in (a) and (b), respectively. All graphs depict spin noise difference spectra.}\label{fig:rb87}
\end{figure*}
Figure~\ref{aufbau}~(b) depicts the experimental setup. Linear polarized laser light is transmitted through a rubidium vapor cell and acquires a stochastic Faraday rotation. The time-dependent Faraday angle is measured by a polarization bridge consisting of a polarizing beam splitter (PBS) and a balanced photoreceiver. The time-domain Faraday rotation data is analyzed via fast Fourier transformation (FFT) spectral analysis. The light source is an external cavity diode laser (ECDL) with a grating for wavelength tuning in Littrow configuration. The laser emission is linear polarized and has a spectral line width of less than 20 MHz. An anamorphotic prism pair yields a round beam shape and an optical Faraday isolator prevents unwanted back-reflections into the laser. The laser beam is focused by a 500 mm focal length lens to a beam diameter of 240~$\mu\mathrm{m}$  at the center of the probed rubidium vapor cell. The Rayleigh range thereby corresponds to the length of the cell of $l=50~\rm mm$. The laser is swept with a tuning range of -7 to +13 GHz relative to the  D2 $F=2\rightarrow F'=3$ transition of $^{87}$Rb at around 780 nm while the absorption of the atomic  system is measured simultaneously. The spectral position of the laser is monitored with a Fabry-P\'erot interferometer. A small magnetic field is applied transverse to the direction of light propagation by means of a Helmholtz coil with the light polarization exactly perpendicular to the magnetic field axis. In this work, two different rubidium vapor cells are investigated: \textit{cell A} with enhanced  $^{87}$Rb concentration and \textit{cell B} with natural abundance. Cell A also contains helium buffer gas at a pressure of 1~mbar. Natural rubidium comprises the different isotopes  $^{85}$Rb (72.15\%) and  $^{87}$Rb (27.85\%). The cells are heated to around 350~K.

The measured quantity in noise spectroscopy is the spectral noise power density, i.e., the amplitude variance per frequency. Subtraction of two acquired noise power spectra from each other remove the spectral noise background which itself results from strong white photon shot noise and residual electrical noise. Figure \ref{example} depicts an typical spin noise spectrum that is acquired with cell A for the case of strong optical detuning. The spin noise peak at 12.8~MHz corresponds to the Larmor frequency for an applied magnetic field of 1.8~mT ($g_F=1/2$). The negative peak close to zero frequency results from the subtraction of a noise power spectrum with no magnetic field applied, i.e., a spectrum where the spin noise signal is centered around zero frequency. The spin noise signal centered at zero frequency shows twice the amplitude compared to the modulated spin noise signal since the spin noise power which would theoretically appear in the second quadrant is folded into the detection window, i.e., for both cases the integrated spin noise power is the same. In the case of strongly broadened spin noise spectra (cell B), the available magnetic field range is not affected by overlapping spectra. Instead, reference spectra are acquired by switching the polarization state of the probe light via a liquid crystal retarder (LCR). This method is described in detail in Ref.~\cite{roemer:apl:94:112105:2009, mueller:physe:2010}. The extracted FWHM of the spin noise peak in Fig.~\ref{example} amounts to $\nu_{\text{FWHM}}=515$~kHz, the corresponding spin dephasing time of $\tau_{s}=620$~ns results from the finite interaction time between the rubidium spins and the probe laser due to the spatial motion of the atoms while the intrinsic spin dephasing time of rubidium is of the order of 100 ms \cite{FranzPRA1976}. Please note that the helium buffer gas of 1~mbar (cell~A) reduces the efficiency of this time-of-flight broadening only slightly and in cell B a corresponding FWHM of around $\nu_{\text{FWHM}}=640$~kHz is measured. Nevertheless, saturation broadening in cell A is very efficient due to the helium buffer gas \cite{Demtroder2003} and the dominating mechanism for broadening of the D2 transitions changes from inhomogeneous Doppler broadening at low probe laser intensities to the homogeneous saturation broadening at high probe laser intensities.

\paragraph*{Spin noise in cell A.}
Next, we study the spin noise in cell A in dependence on the laser detuning and the probe laser intensity. Figures \ref{fig:rb87} (a) and (b) depict the detected spin noise power density in color coding as a function of the laser detuning (abscissa) and the noise frequency (ordinate) at probe laser intensities of $I_{l}=1.8$ and $14.1\,\mathrm{W/cm^2}$, respectively. The black line depicts the corresponding absorption spectra. The intensity plots show the difference noise spectra between applied external magnetic fields of $\approx 0$ and $1.8\, \mathrm{mT}$, respectively, i.e., different Larmor frequencies centered at $\approx 0$ and $12.8\,\mathrm{MHz}$. The overall absorption spectrum at low probe intensities of the $F=2\rightarrow F'=1,2,3$ transitions show a FWHM on the order of 2~GHz where the Doppler broadening contributes by an inhomogeneous line width of 560~MHz (calculated value). The spin noise power density resulting from the $F=2\rightarrow F'=1,2,3$  transitions around 0~GHz detuning shows a pronounced asymmetric dependency for positive and negative detuning. The asymmetry of the spin noise power spectrum decreases with increasing laser intensity [compare Fig.~\ref{fig:rb87} (a) and (b)] since the $F=2\rightarrow F'=1,2,3$ transitions are simultaneously excited at high intensities due to the stronger homogeneous saturation broadening.
We will show in Section \ref{theory} that N-type coherences are responsible for the asymmetric behavior and the spectral shift between absorption maximum and spin noise minimum.

\paragraph*{Spin noise in cell B.}
\begin{figure}[ht]
        \includegraphics[width=\columnwidth]{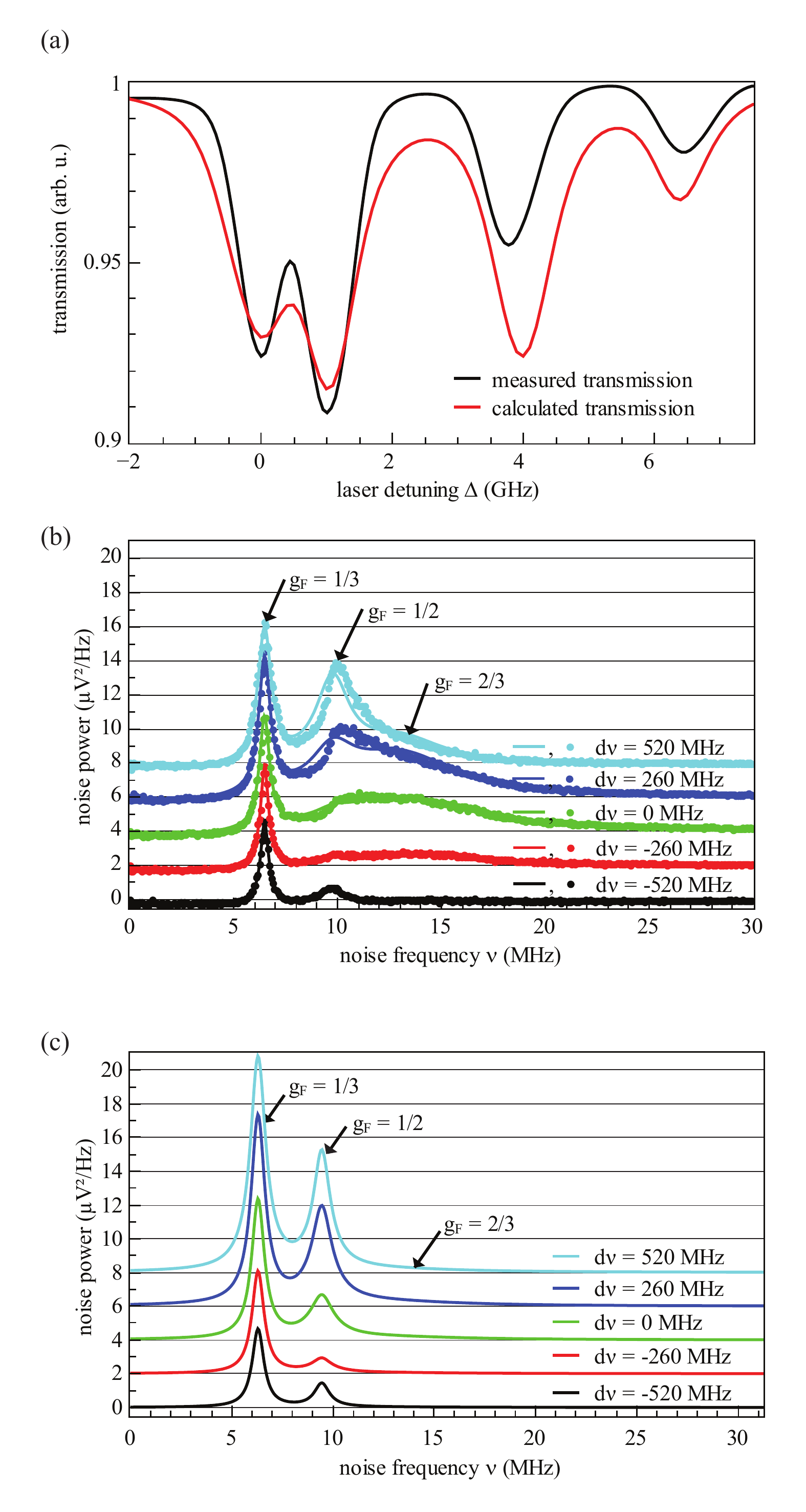}
    \caption{(Color online) (a) Calculated and measured absorption spectra for cell B (natural rubidium). (b) Measured spin noise spectra of cell B with a probe laser intensity of $I_{l}=7.8\,\mathrm{W/cm^2}$ for different laser frequencies. At low detuning, spin noise of the excited $5^2P_{3/2}$ states ($g_F=2/3$) becomes significant. (c) Spin noise spectra calculated by the extended Bloch model corresponding to the experimental data shown in (b).}\label{fig:rbnat}
\end{figure}
Figure ~\ref{fig:rbnat} (a) and (b) depict absorption and spin noise spectra for cell B acquired with a probe laser intensity of $I_{l}=7.83\,\mathrm{W/cm^2}$. The absorption spectra are dominantly broadened by saturation broadening. The spin noise spectra exhibit an additional noise signal due to the different Land\'e factor of the ground state of $^{85}$Rb ($g_F=1/3$) compared to $^{87}$Rb. Most strikingly, the excited $5^2P_{3/2}$-$^{87}$Rb states also contribute to the observed spin noise at high probe intensities and low detuning from the optical resonances. This is indicated at the corresponding Land{\'e}-g-factor of $g_F=2/3$ in Fig.~\ref{fig:rbnat}~(b). The excited state $g_F$-factors of $^{85}$Rb are $-1$,  $1/9$, $7/18$ and $1/2$ for $F'=1,2,3,4$ and hence the excited state spin noise of $^{85}$Rb is spread over a huge frequency interval and thus does not interfere with the measured excited state noise of $^{87}$Rb.


\section{Theory}\label{theory}
\paragraph*{Bloch equations.}
The detected spin noise results from a circular birefringence due to the fluctuating occupation of the degenerate $m_F$ spin states \footnote{note1}. The equilibrium occupation probability of each $S$- and $P$-type hyperfine level $F$ is needed to determine the fluctuation strength and the transition probability of each transition involved and henceforth to calculate the circular birefringence. All necessary quantities are obtained by solving the optical Bloch equations. The optical Bloch equations represent the equation of motion for the density matrix ${\rho }$ of the investigated atomic system in the presence of a light field \cite{ArimondoNCL1976, RenzoniPRA1997}. In the following, the index $i$ ($j$) denotes the ground (excited) states. The ground states for the case of $^{85}$Rb ($^{87}$Rb) are  $5^2S_{1/2}$  $F=2,3$ ($F=1,2$); the excited states  are  $5^2P_{3/2}$  $F'=1,2,3,4$ ($F'=0,1,2,3$). In the rotating wave approximation, the Bloch equations for the ground state occupancies $\rho _{ii}$ read \footnote{note2}:
\begin{eqnarray}
\dot{\rho}_{ii} &=& \sum_{j} \gamma_{ij} \cdot \rho_{jj} + 2 \sum_{j} \Omega_{ij} \cdot \rm{Im} \{\rho_{\it{ij}}\}  \nonumber \\
 && \;+   \gamma_{\rm{diff}} \sum_{i'}\varepsilon_{ii'} \cdot \rho_{i'i'}   \frac{g_{i}}{\sum_{i} g_{i}} .
\end{eqnarray}
Here, $\gamma _{ij} ={2\pi \, \nu_{ij}^{2} e^{2} }({m\, c^{3} \varepsilon_{0} })^{-1} {f_{ij} }/({f_{1j} +f_{2j} }) f_{JJ'}$ is the spontaneous transition rate where $f_{JJ'} =0.6958$ denotes the fine structure oscillator strength, the term $f_{ij} $ yields the branching ratio of the hyperfine transition with
\[
f_{ij} = \left( \begin{array}{cccc} {1/2} & {5/4} & {5/4} & {0} \\{0} & {1/4} & {5/4} & {7/2} \end{array} \right),
\]
and $\nu_{ij}$ is the resonance frequency of the corresponding optical transition.
 The resonant Rabi frequency for  linear polarized light  with intensity $I_{l}=c\epsilon_0 E_0^2/2$ and light frequency $\omega _{ij}$ reads
\[
\Omega _{ij} = \frac{\mu_{ij} \cdot E_0} {2 \hbar } =\frac{1}{2}\sqrt{\frac{e^{2} }{m\, \hbar \, \omega _{ij} } \frac{I}{c\, \varepsilon _{0} } \frac{f_{ij} }{g_i } f_{JJ'} }.
\]
Here, $E_{0}$ is the electric field amplitude of the probe laser light and $\mu_{ij}$ the transition dipole moment. Relaxation towards the thermal equilibrium by diffusion of atoms in and out of the excitation area is accounted by the diffusion rate $\gamma _{\mathrm{diff}} = \pi \cdot \nu_{\mathrm{FWHM}}$, which is extracted from the off-resonant noise spectra and amounts to $\pi\cdot 515\,\mathrm{kHz}$ for cell A and $\pi\cdot  640\,\mathrm{kHz}$ for cell B. The occupation probability of the valence electron in the corresponding ground state scales with the degeneracy $g_{i}=2i +1$ and $\varepsilon _{ii'}$ is the two dimensional permutation tensor.
The excited state occupancies are given by
\begin{equation}
\dot{\rho }_{jj} =-\sum _{i} \left[ \gamma _{ij} \rho _{jj} + 2 \, \Omega _{ij} \cdot \rm{Im}\{\rho_{\it{ij}}\}\right] . 
\end{equation}

The coherences between ground and excited states are given by the off-diagonal elements of the density matrix and have to be included to correctly model the experimental findings, such as the asymmetric shape of the spin noise spectrum versus detuning:
\begin{eqnarray}
\dot{\rho }_{ij} &=&\left(-\frac{1}{2} (\gamma_{ij} +\gamma_{B} )-i\Delta _{ij} \right)\cdot \rho_{ij} +i\cdot \Omega_{ij} \cdot (\rho_{jj} -\rho _{ii} ) \nonumber \\
&& +i\sum_{j', i-j'=\pm 1}\Omega_{ij'} \cdot \rho_{jj'}  + i\sum_{i', i'-j=\pm 1}\Omega_{i'j} \cdot \rho_{ii'}\, . \label{eq:g-e-coh}
\end{eqnarray} 
Here the additional relaxation rate $\gamma_{B} $ describes the quenching of the excited states after collisions with  buffer gas atoms. For 1~mbar helium gas pressure (cell A), $\gamma_{B}$ is extrapolated to $2\pi \cdot 18\,$MHz from the data presented in Ref~\cite{PhysRevA.56.4569}. The detuning is given by $\Delta_{ij} =\omega_{ij} - \omega'$ with $\omega'=\omega_{l} (1- u/c) $ for atoms with a velocity $u$, the laser frequency $\omega_l$, and resonance frequencies $\omega_{ij}=2\pi\,\nu_{ij}$ \cite{Steck2009a,Steck2009}.

Note that in Eqs. (\ref{eq:g-e-coh}) the ground-ground coherences ($\rho_{ii'}$) and the excited state coherences ($\rho_{jj'}$) are included, and hence the dynamics of these coherences must be carefully considered:
\begin{eqnarray}\label{eq:GroundStateCoherence}
\dot{\rho}_{ii'} &=&i \sum_{j} [ \delta_{j,i \pm 1,0} \cdot \Delta_{ij}  -\delta_{j,i'\pm 1,0} \cdot \Delta_{i'j} ] \cdot  \rho_{ii'} \nonumber \\ &&+ i \sum_{j} [\Omega_{ij} \cdot \rho_{i'j} - \Omega_{i'j} \cdot \rho_{ji} ],
\end{eqnarray}
\begin{widetext}
\begin{equation}\label{eq:ExcitedStateCoherences}
\dot{\rho }_{jj'} =\left\{ -\frac{1}{2} \left( \sum _{i} [\gamma _{i j} + \gamma_{i j'}]  \right) -\gamma _{B} +i\sum_{i} [\delta_{j,i\pm {\rm 1,0}} \cdot \Delta_{i\pm {\rm 1,0\; }j} -\delta_{j',i\pm {\rm 1,0}} \cdot \Delta_{i\pm {\rm 1,0\; }j'} ] \right\}  \cdot \rho_{jj'} 
 + i\sum _{i}[\Omega_{ij} \cdot \rho _{ij'} -\Omega_{ij'} \cdot \rho _{ji} ]  , 
\end{equation}
\end{widetext}
where $\delta _{j,i\pm {\rm 1,0}}$ ensures the $\Delta F=0,\pm 1$ selection rules.
 
The steady state of this system of differential equations is solved numerically in terms of $\omega'$.  The Doppler broadening is included by convolution of the respective quantity of interest with the velocity distribution $f(u)=({\sqrt{2\pi } \bar{u}})^{-1} e^{-(u/\bar{u})^{2}/2 }$ of atoms with the velocity $u$ and an average velocity $\bar{u}=\sqrt{k_{B} T / m}$ for a given temperature $T$. Note that every density matrix element $\rho$ is a function of $u$.

\paragraph*{Absorption spectrum.}
The absorption spectrum is calculated from the system's susceptibility $\chi (\omega')$ including all $\sigma_{(at)}$- and $\pi_{(at)}$-type transitions: 
\begin{equation}\chi  (\omega')
  =\frac{n_{A} }{\varepsilon_{0} E_{0} } \sqrt{\frac{\hbar \, e^{2} }{2\, m} } \sum _{ij,\Delta m_{F} =0,\pm 1}\sqrt{\frac{f_{ij} }{\omega _{ij} } } \rho _{ij} (\omega'), 
\end{equation}
with $n_A$ being the rubidium atom density. The Doppler broadening is included by convolution of $\chi  (\omega')$ with the velocity distribution $f(u)$. The absorption coefficient $\kappa$ is calculated by $\kappa  (\omega')=\rm{Im} \{ \sqrt{1+\chi  (\omega')} \}$.

\paragraph*{Spin noise spectrum.}
Calculating the spin noise spectrum is not as straight forward as the absorption spectrum since each hyperfine split optical transition $F\rightarrow F'$ needs to be weighted with the mean square spin imbalance $\langle m_F^2 \rangle$, i.e., the actual spin noise, and the spin noise of the distinct hyperfine levels $F$ appear at the corresponding Larmor frequency $\nu_{\mathrm{L}}=g_F\mu_{\mathrm{B}}B/h$ in the noise spectrum. 
Correspondingly, we model the complete spin noise spectrum as a sum of Lorentz functions \footnote{note3} for all states. The ground state spin relaxation rates increase at zero detuning due to optical pumping by:
\begin{equation}\label{eq:NoiseWidth}
\Gamma_i=2\left(\gamma_{\mathrm{diff}}+\frac{2}{3}\left[\sum_j\gamma_{ij}\rho_{jj} +\sum_{i'\neq i,j} \gamma_{i'j}\rho_{jj}  \right] \right). 
\end{equation}
The noise amplitude is given by:
\begin{equation}\label{eq:NoiseAmplitude}
U_i=\alpha\, P_{l} \sin(d\theta_i)\cdot e^{-\kappa \, l\, k_z}. 
\end{equation}
Here $\alpha=20\, \rm V/\rm W$ is the amplification factor of the balanced receiver, $P_{l}$ the laser power, and $k_z$ the wave vector.
The Faraday noise angle is calculated by \footnote{note4}:
\begin{eqnarray}\label{eq:NoiseAngle}
d\theta_i^2 &=& \left( k_z l \frac{\Delta n_i}{2} \right)^2 \\
 &=& \left(\frac{k_z l\, \mu_{i j}}{\varepsilon_0 E_0 3} \right)^2 \left(\frac{\sqrt{N}}{V} \right)^2 \, \mathrm{Re} \{\rho_{ij}\}^2\rho_{ii} . \nonumber
\end{eqnarray}
The sum of Lorentz functions is finally convoluted with the velocity distribution to include Doppler broadening and is valid for uncorrelated atoms \footnote{note5}. In Eq.~\eqref{eq:NoiseAngle} $\mu_{i j}/{3}$ is the dipole matrix element for each $\sigma^{\pm}$ transition which is calculated separately for every hyperfine transition. The corresponding expressions for the excited states are obtained by interchanging $i$ and $j$ in Eqs.~(\ref{eq:NoiseWidth},\ref{eq:NoiseAmplitude},\ref{eq:NoiseAngle}). The excited state spin relaxation rates are calculated by $\Gamma_j=2(\gamma_{\rm{diff}}+\frac{2}{3}\sum_i (\gamma_{ij}+\rho_{jj} \gamma_{ij}))$. Further a factor of two is included in the calculation of the noise power of the excited states to account for stimulated emission.

\paragraph*{Results and comparison with experimental data.}
Calculated absorption and spin noise spectra  are depicted in Figs.~\ref{fig:rb87}~(c,d) and \ref{fig:rbnat}~(a,c).  The calculated absorption spectra reproduce qualitatively, and to a large extent quantitatively, all different broadening mechanisms. The saturation broadened absorption spectra show a non-negligible deviation in the width of the corresponding transitions between theory and experiment. This discrepancy results from the probe laser beam waist that varies in the experiment over the cell length with a factor of $\sqrt{2}$. However, this inaccuracy in the calculations is negligible since the spin noise at the focus point is weighted stronger than the spin noise at the cell endings. Indeed, the calculated and measured spin noise spectra for cell A (pure $^{87}$Rb, Fig. \ref{fig:rb87}) are in excellent agreement. Please note that there is no free parameter in the calculations to adjust the absolute noise power values. Especially, the asymmetric behavior around the $F=2\rightarrow F'=1,2,3$ transition is remarkably well reproduced for the two different noise spectra. This asymmetry decreases in the calculations as well as in the experiment for higher probe light intensities due to additional homogeneous broadening. The ratio between laser intensity and spin noise signal stays constant due to the larger homogeneous broadening and increasing spin relaxation rates for stronger laser light fields.

In the case of cell B (Rb with natural abundance), the spin noise arising from the excited $5^{2}P_{3/2}$-$^{87}$Rb states is qualitatively also well reproduced [Fig.~\ref{fig:rbnat} (b,c)]. The excited state noise corresponds to noise frequencies around $15$ MHz denoted by $g_F=2/3$. We note, however, one significant discrepancy between the experimental and theoretical results in what concerns the strength of the excited state spin noise. The enhanced excited state spin noise measured at detunings close to the optical resonance and high probe laser intensities is not recovered by the single-particle Bloch analysis. This discrepancy indicates a signature of a collective effect such as nonlinear self-rotation \cite{BudkerRMP2002} and polarization selective absorption of the probe light while traversing the rubidium gas cell. These many-body effects are beyond the scope of the single particle model discussed in this paper and only appear at significant occupation of the excited state.


\section{Conclusion}
We have performed spin noise spectroscopy on a classical rubidium gas under resonant and non-resonant probing conditions. The experimental results are modeled by extended Bloch equations, which inherently include the populations and coherences between all participating states. Our careful analysis shows that the coherences between ground and excited states, and also ground-ground and excited-excited coherences, play a crucial role in spin noise spectroscopy at resonant and quasi-resonant probing conditions. This is contrary to the case of off-resonant probing where coherences play no significant role. We identify a characteristic asymmetric spin noise signal with respect to detuning which arises due the coherent coupling of the atomic levels under nearly resonant probing conditions. The theoretical results are in very good qualitative, and to a large extent quantitative, agreement with our experimental findings. However, a significant deviation is observed for the specific case of excited state spin noise for nearly resonant probing conditions. Interestingly, this discrepancy may indicate the appearance of collective effects, which are not covered by single-particle extended Bloch equations. We expect that these and similar effects also plays an important role in spin noise experiments with self-assembled semiconductor quantum dots or donor atoms.

\section{Acknowledgments}
We gratefully acknowledge financial support by the BMBF \textit{QuaHL-Rep}, the Deutsche Forschungsgemeinschaft in the framework of the priority program SPP1285 \textit{Semiconductor Spintronics}, and by the German Excellence Initiative via QUEST - \textit{Center for Quantum Engineering and Space-Time Research}.

\bibliographystyle{apsrev4-1}   
%

\end{document}